\documentclass[prd,showpacs,preprintnumbers,floatfix,superscriptaddress,10pt]{revtex4}
\usepackage[mathscr]{eucal}
\usepackage[dvips]{color}
\usepackage[dvips]{graphicx}
\usepackage{epsf}
\usepackage{bm}
\usepackage{amssymb}
\usepackage{amsmath}
\usepackage[normalem]{ulem}
\newcommand{\be}{\begin{equation}}

\newcommand{\ee}{\end{equation}}
\newcommand{\ba}{\begin{eqnarray}}
\newcommand{\ea}{\end{eqnarray}}

\newcommand{\bea}{\begin{eqnarray}}
\newcommand{\eea}{\end{eqnarray}}
\begin{document}

\title{R-mode oscillations and rocket effect  in  rotating superfluid neutron
stars. \\I. Formalism}
\author{Giuseppe~Colucci}
\affiliation{Universit\`a di
Bari, I-70126 Bari, Italy} \affiliation{I.N.F.N., Sezione di Bari,
I-70126 Bari, Italy}
\author{Massimo~Mannarelli}
\affiliation{Departament d'Estructura i Constituents de la Mat\`eria and
Institut de Ci\`encies del Cosmos (ICCUB), Universitat de Barcelona,
Mart\'i i Franqu\`es 1, 08028 Barcelona, Spain\\ I.N.F.N., Laboratori Nazionali del Gran Sasso, Assergi (AQ), Italy}
\author{Cristina~Manuel}
\affiliation{Instituto de Ciencias del Espacio (IEEC/CSIC),
Campus Universitat Aut\`onoma de Barcelona, Facultat de Ci\`encies, Torre C5
E-08193 Bellaterra (Barcelona), Spain}
\begin{abstract}
We derive the hydrodynamical equations of r-mode oscillations in neutron stars in presence of 
a novel damping mechanism related to particle number changing processes.  The change in the number densities of the various
species  leads to new dissipative
terms in the  equations which are responsible of the {\it rocket effect}. 
 We employ a two-fluid model, with one fluid consisting of the charged components,
 while the second fluid consists of superfluid neutrons. 
We consider two different kind of r-mode oscillations, one associated with comoving displacements, 
and the second one  associated with countermoving, out of phase, displacements. 
\end{abstract}
\preprint{BA-TH/629-10}
\date{\today}
\maketitle
\section{Introduction}
Rapidly rotating neutron stars  have  been the
subject of intensive investigation in the last years. Of particular
interest  are  neutron star oscillations, which might be useful
to shed light on the internal structure of these
stars~\cite{Andersson:2000mf,Lindblom:2001}. Stars have  various
modes of oscillations,  among them  r-mode
oscillations are probably the most interesting ones, because
they provide a severe limitation on the star's rotation frequency through
coupling to gravitational radiation  emission~\cite{Andersson:1997xt,Friedman:1997uh}.
The  oscillations  of compact stars can be damped by  various dissipative
process~\cite{Lindblom:1998wf, Andersson:1998ze}, which take place in the interior of the star. 
However, if dissipative phenomena are not strong enough,  the r-mode oscillations will grow 
exponentially fast in time until the star slows down, by emission of gravitational waves, 
to a rotation frequency where some dissipative mechanism efficiently damps these oscillations. 
Since neutron stars are observed to rotate at very high frequencies, any model of neutron star 
must provide an efficient mechanism of dissipation of r-mode oscillations. In this way, the study of
r-mode damping  is useful in  constraining  the stellar structure and can be
used to rule out some exotic phases of matter~\cite{exotic1,exotic2,exotic3}. 

Standard neutron stars are stellar objects  with a mass of about $1.4 M_{\odot}$
and a radius of about $10$ Km. They are believed to have
a crust of about $1$ Km, with an outer part made of a lattice of ions embedded
in a liquid of electrons and an inner part made of nuclei embedded in a liquid
of $^1S_0$ superfluid neutrons.  In the interior of the star nuclei are melted and both neutrons and protons 
are expected to condense into BCS-like superfluids. However, neutron interaction
in the $^1S_0$ state at supernuclear matter density is repulsive, but it is
still possible to form
Cooper pairs in the $^3P_2$ channel~\cite{Hoffberg:1970}. The proton density is
much smaller than the neutron density and therefore the formation of $pp$ Cooper pairs  in
the isotropic $^1S_0$ channel is allowed.
Pairing between protons and neutrons does not take place because of the large
mismatch between their Fermi energies. 
In the core of neutron stars muons might be present (when $\mu_e > m_\mu$), or  
deconfined quarks  in a color superconducting phase, 
moreover  pion or kaon condensates might be realized. In the present paper
we shall not consider any of these possibilities and assume the core of the
neutron star comprises only neutrons, protons and electrons.

R-mode oscillations have been studied extensively in the literature~\cite{Andersson:2000mf}, 
and it is known that shear and bulk viscosities are able to suppress the instability in two different ranges of
temperatures~\cite{Lindblom:1998wf, Andersson:1998ze}.   For temperatures
smaller than about $10^5$ K the fluid damping due to shear viscosity suffices to suppress the r-mode instability, but with
increasing temperature the effect of shear viscosity is gradually suppressed.
On the other hand,  for temperatures larger than about $10^{10}$ K -$10^{11}$ K, bulk viscosity becomes 
an efficient mechanism for damping  r-mode oscillations. Bulk viscosity does not lead to sufficient 
damping at lower temperatures because neutron matter is likely to be in the superfluid phase where bulk viscosity is
suppressed  by Pauli blocking. However, at temperatures above  $10^{10}$ K -$10^{11}$ K 
 nuclear matter is believed to be in the normal phase  with a large bulk viscosity coefficient.  
 Therefore, one has an ``instability window" for standard neutron stars corresponding to  a
range of temperature approximately given by  $10^6$\,K~-~$10^{10}$\,K. The exact
values depend on the details of the model considered and on the rotation frequency of the star. 
The instability window is in part reduced  by the ``surface rubbing" between the
core and the crust of the star~\cite{Bildsten:1999zn,Glampedakis:2006mn}. This  
mechanism results in a viscous boundary layer between the core and the crust of
the star  which damps r-mode oscillations for temperatures less than about
$10^{10}$ K and for sufficiently small frequencies.

In Refs.~\cite{Lindblom:1999wi, Lee:2002fp, Andersson:2006, Haskell:2009fz} it 
is studied the effect of mutual friction  in reducing the instability window. It
is shown that the typical timescale of mutual friction is of the order of $10^4$s 
and is therefore too long for damping the r-mode instability. Indeed, the 
timescale associated with gravitational-wave emission is of the order of few
seconds (for a millisecond pulsar). However, mutual friction can reduce the
instability for certain values of the entrainment
parameter~\cite{Lindblom:1999wi} or for large values of the drag
parameter~\cite{Haskell:2009fz}. 

In the present paper we determine the hydrodynamical equations for r-mode oscillations 
in presence of a novel dissipative mechanism associated with the change in the number of protons,
neutrons and electrons which we shall refer to as  the {\it rocket effect}. In real neutron stars 
this mechanism can take place  in the outer core and in the inner crust of the star and is related
to beta decays and interactions between the neutron fluid and the crust. 
The rocket effect is dissipative because when two or more fluids move with different
velocities a  change of one component into the other results in a 
 momentum transfer between the fluids. This change in momentum is not reversible,
because it is always the faster moving fluid that will lose momentum.  The resulting 
dissipative force is  proportional to the mass rate change, and to the relative velocity between the  fluids.
The name ``rocket effect" reminds that the same phenomenon  takes place  
in the dynamical evolution of a rocket whose mass is changing in time as it
consumes its fuel. 

In order to simplify the analysis we consider a simplified model of neutron star
consisting of a fluid of neutrons, protons and electrons and no crust. Protons and 
electrons are locked together by the electromagnetic interaction and therefore we 
consider that the system consists of two fluids. As a further simplification we
assume that the star has a uniform mass distribution with density $\rho = 2.5
\rho_0$ and a radius of $10$ km.  Since this simplified model of star does not
comprise a crust we consider only number changing processes associated with weak interactions.  

In our analysis we consider two different r-mode oscillations.
One is associated  predominantly with toroidal comoving displacements and the second one is dominated by  toroidal countermoving displacements. We shall
refer to these oscillations as ``standard r-mode" and as
``superfluid r-mode", respectively. These two modes decouple
for a star made by  uniform and incompressible matter, and we shall restrict to treat  such a case.  
By performing an expansion in the parameter $\Omega/\Omega_K$, where $\Omega$ is the frequency of the star
and $\Omega_K$ is its Kepler frequency, we find that the linearized equations  for both the standard and superfluid
r-modes present additional dissipative terms due to the rocket effect, but for each oscillation they appear in a 
different order in our expansion parameter. The numerical evaluation of the timescale related to this mechanism is 
performed in the accompanying paper \cite{rocket2:2010}.

This paper is organized as follows. In Section \ref{sec-hydro} we review the
hydrodynamic equations in the two fluid approximation. 
In the hydrodynamical equations we neglect shear and bulk viscosities but  consider
 the mutual friction force and the rocket effect. 
In Section \ref{sec-perturbed} we present the differential equations governing
the deviations from equilibrium of a rotating two fluid system. We study the
linearized problem,  neglecting deformations of the star due to rotation  and
 assuming uniform mass density. Although not very realistic, these
assumptions allow us to simplify the study of the r-modes, obtaining an a analytical 
expression for the fluid displacements. We draw our preliminary conclusions in Section \ref{sec-conclusion}.
In Appendix \ref{sec-evolution} we report some details about the evolution
equations for the comoving and countermoving displacements. 

\section{Hydrodynamical equations for the multifluid neutron star
model}\label{sec-hydro}

The equations describing the dissipative processes of neutrons, protons and
electrons in the outer core of a standard neutron star have been studied in
detail in Ref.~\cite{Andersson:2006}. In general,  the entropy production
rate depends on 19 independent coefficients which are related to the various
dissipative processes.  However, for low temperatures, well below the critical temperature for superfluidity, and
neglecting viscosities one obtains the expression
given in Ref.~\cite{Prix:2002jn} which depends only on two different
coefficients.
One is related to mutual friction and the second one with  the so-called
rocket term.
Here we review the basic hydrodynamical equations in presence of 
these two different dissipative
mechanisms, neglecting the presence of  shear and bulk viscosities.

For a system consisting of 
neutrons,  protons  and  electrons,
the mass conservation law is given by, see {\it
e.g.}~\cite{Prix:2002jn,Andersson:2006},
\begin{equation}\label{continuity}
 \partial_t\rho_x+\nabla_i(\rho_x v_x^i)= \Gamma_x \ ,
\end{equation}
where $\Gamma_x$ is the particle mass creation rate per unit volume and the 
index $x=n,p,e$ refer to the particle species, that is, neutrons, protons and
electrons. In these equations we have  considered that some process  can
convert neutrons in protons and electrons and {\it vice versa}.
Therefore, we are assuming that the various components are not separately
conserved. One possible mechanism leading to a change in 
the particle number densities is  given by the weak processes 
\begin{equation}
n  \rightarrow  p + e^- + {\bar \nu}_e \, , 
\qquad
p + e^- \rightarrow n + \nu_e \,.
\end{equation}
These  reactions  lead to a  change in  the chemical potentials of the
various species and therefore are associated with  number density changes.

A different process can lead to a non-vanishing mass creation rate, which we shall call {\it crust-core transfusion}.
In this process when a compression takes place, the ionic
constituent of the crust are squeezed and part of their nucleonic content is
released and
augments the fluid components.  The opposite mechanism, related to
a reduction of the pressure leads to the nucleonic capture by the ions of the
crust. In the present section we  consider that a generic mechanism is at
work to produce a change in the number densities. In the accompanying paper \cite{rocket2:2010}
we shall evaluate the particle mass creation rate corresponding to the beta decay
processes. Regarding the crust-core transfusion processes the corresponding creation
rates are difficult to evaluate and we postpone their calculations to future
work. 

In any case, the three particle creation rates are not independent quantities,
because charge conservation implies that
\be
\frac{\Gamma_e}{m_e} = \frac{\Gamma_p}{m_p} \, ,
\ee
whereas baryon number conservation leads to
\be
\frac{\Gamma_p}{m_p} = -\frac{\Gamma_n}{m_n} \, ,
\ee
meaning that only one creation rate is independent.

It is possible to simplify the treatment of the system considering that 
our analysis regards processes that happen at a time scale
much larger than the electromagnetic time scale. 
Therefore, we can consider that  electrons and protons are locked together to
move with the same velocity~\cite{Alpar:1984}, see
however~\cite{Mendell:1991}. Moreover, charge neutrality implies
that the number densities of electrons and protons are equal, {\it i.e.} $n_e =
n_p$, meaning that electrons and protons can be treated as a single charge-neutral fluid
and henceforth we shall refer to this fluid as the ``charged'' component,
employing the subscript $c$ to label it. As a matter of fact, the system can be
viewed as consisting of two fluids, with mass densities 
\be
\rho_n = m_n\, n_n \qquad {\rm and } \qquad \rho_c = m_n\, n_c \,, 
\ee
where $m_n=m_p + m_e$ and $n_e=n_p=n_c$.  For non vanishing mass creation rates, the Euler equations have an extra
term, see~\cite{Prix:2002jn}, and are given by 
\begin{eqnarray}
\label{Eul-neutron}
 (\partial_t +v_n^j\nabla_j)(v_i^n+\epsilon_n
w_i)+\nabla_i(\tilde{\mu}_n+\Phi)+\epsilon_n w^j\nabla_iv_j^n
& = & \frac{f^{\rm MF}_i}{\rho_n} \,,\\
\label{Eul-proton}
(\partial_t +v_c^j\nabla_j)(v_i^c - \epsilon_c w_i)+\nabla_i(\tilde{\mu}_c+\Phi)
-\epsilon_c w^j\nabla_iv_j^c & = &- \frac{f^{\rm MF}_{i}}{\rho_c} + \left(1
-\epsilon_n -\epsilon_c  \right) \frac{\Gamma_n}{\rho_c} w_i \,,
\end{eqnarray}
where $i,j$ label the space components,  we have defined a chemical potential
by mass $\tilde{\mu}_x = \mu_x/m_n$,  and 
 ${\bf w} = {\bf v}_c -{\bf v}_n$ represents the relative velocity between the
two fluids.  The quantities   $\epsilon_n$ and $ \epsilon_c$ are 
the entrainment parameters, that are related to the fact that momenta and
velocities of quasiparticles may not be aligned~\cite{Andreev}, and the the gravitational potential,
$\Phi$, obeys  the Poisson
equation
 \begin{equation}
 \nabla^2 \Phi = 4 \pi G \left( \rho_n + \rho_c \right) \,.
 \end{equation}

The force term $f^{\rm MF}_i$ entering into both Euler equations corresponds to
the mutual friction  force between the  superfluid and normal component.  This force appears when a superfluid is put
in rotation~\cite{HallV,Sonin:1987zz,knopin} and  at the microscopic level
it is due to  the scattering  of  the normal component  off the superfluid vortices~\cite{Alpar:1984}. In the present
case it is due to the scattering of electrons off the neutron superfluid  vortices.  Indeed, as a consequence of the
entrainment between neutrons and protons the superfluid vortices are accompanied by a magnetic field. The expression of
the mutual friction force valid for small values of $w$ has been determined in~\cite{HallV} and is  given by
\begin{equation}\label{force-mf}
 f^{\rm MF}_i=
2\rho_n {B^\prime} \epsilon_{ijk}\Omega_jw^k+2\rho_n B \epsilon_{
ijk }\hat{\Omega}^j\epsilon^{klm}\Omega_lw_m \,,
\end{equation}
where the coefficients $B, B^\prime$ can be written as
\be
B= \frac{{\cal R}}{1+{\cal R}^2}, \qquad {\rm and}  \qquad B^\prime =
\frac{{\cal R}^2}{1+{\cal R}^2} \,,
\ee
where ${\cal R}$ is the dimensionless ``drag'' parameter~\cite{Andersson:2006}. 
The actual strength of the drag is not precisely known, see {\it
e.g.}~\cite{Alpar:1984,Ruderman:1998}, and  one can consider three different regimes: the weak
drag regime, ${\cal R} \ll 1$, the strong drag regime ${\cal R} \gg 1$ and
the intermediate drag regime, ${\cal R}\sim 1$.  For small values of the drag
parameter one can express the coefficients $B, B^\prime$ as a function of the
entrainment parameter. Considering scattering of electrons off vortices, according with Ref.~\cite{Alpar:1984},
one has that 
\be\label{B-epsilon}
B = 4 \times 10^{-4} \frac{\epsilon_c^2}{\sqrt{1-\epsilon_c}}\left(
\frac{x_c}{0.05}\right)^{7/6}\left(\frac{\rho}{10^{14} {\rm
g/cm}^3}\right)^{1/6} \qquad {\rm and} \qquad B^\prime \simeq B^2.
\ee

The last term on the right hand side of Eq.~(\ref{Eul-proton}) is the so-called rocket term.  
This force is  due to the fact that  when two fluids  move with different velocities a  change of one 
component into the other results in a variation of the momentum of each fluid component. This change 
in momentum can be viewed as a force  proportional to the mass rate change,
$\Gamma_n$, and to the relative velocity between the two fluids, ${\bf w}$.
Actually, in Eqs.~(\ref{Eul-neutron},\ref{Eul-proton}), one can see that the
rocket term acts only on the charged component.  The reason is that in the presence of the rocket
term, the mutual friction is not uniquely determined because part of the
rocket term contribution can be included in the definition of the mutual
friction force. In the present analysis  we have employed the same definition given  in
Ref.~\cite{Prix:2002jn}. One can show, see Ref.~\cite{Prix:2002jn} for more details, that  the   mutual
friction force is given by the expression in Eq.~(\ref{force-mf}).

In summary, in the presence of the rocket effect one has to  consider a nonvanishing mass
creation rate in  Eq.~(\ref{continuity}), and the rocket term force in Eq.~(\ref{Eul-proton}).
As we shall show in accompanying paper \cite{rocket2:2010},
the rocket effect leads to energy dissipation, and we shall estimate the
corresponding damping timescale for r-mode oscillations. In previous analysis of the possible dissipative
mechanisms of star oscillations
the rocket term has been neglected. Indeed, it was assumed that the neutron, proton and electron numbers are
separately conserved quantities, that is $\Gamma_p = \Gamma_e =\Gamma_n =0$. 

\section{Perturbed  hydrodynamical equations}\label{sec-perturbed}

A non-vanishing mass creation rate influences the evolution of the various hydrodynamical quantities. Indeed, the continuity equation
(\ref{continuity}) as well as the  Euler
equations (\ref{Eul-proton}) depend on $\Gamma_n$. Therefore, in the analysis of the various modes of oscillations of a
neutron star one has to take into account effects related to this term.
In the present paper we   only discuss its  effect on the evolution of the
r-modes of a superfluid neutron star, although it would be equally interesting
to study its effect on other pulsation modes.  In the following analysis of the
hydrodynamical equations we also include
the mutual friction force and we  follow very closely the recent
analysis  of the r-mode oscillations  developed in Ref.~\cite{Saio} for
normal fluid stars, and extended to superfluid stars in
Refs.~\cite{Haskell:2009fz,Andersson:2008fg}.

As in Ref.~\cite{Haskell:2009fz}, we study the linearized hydrodynamical
equations for the perturbations around an equilibrium configuration of a
neutron star rotating with constant angular velocity $\Omega$, 
and we assume that the background configuration is such that the two fluids
move with the same velocity, thus at equilibrium ${\bf w} =0$.  

It is useful to write the Euler equations for the perturbed quantities using
as degrees of freedom (dof) the center of mass displacement and the relative
displacements between the neutron fluid and the charged fluid.  We define the
comoving velocity  as 
\begin{equation}
 \delta {\bf v}=\frac{\rho_n}{\rho}\delta {\bf v_n}+\frac{\rho_c}{\rho}\delta
{\bf v_c} \, ,
\end{equation}
and  the countermoving, or relative, velocity as
\begin{equation}
 \delta {\bf w}=\delta {\bf v_c}-\delta {\bf v_n} \,.
\end{equation}

The continuity equation for the comoving degree of freedom  is not affected
by the rocket effect and is given by
\begin{equation}
 \partial_t\delta\rho+\nabla_j(\rho\delta v^j)=0 \,,
\end{equation}
on the other hand  the continuity equation for the
countermoving dof depends on it. We shall assume that the mass creation rate is given by
\be
\Gamma_n = \bar \Gamma_n + \delta \Gamma_n \,,
\ee
where $\bar \Gamma_n$ is a steady mass creation rate,
and  $\delta \Gamma_n$ is a small fluctuation on the top of it.
Then,  employing as a second  continuity equation  the one for 
the charged fraction, $x_c=\rho_c/\rho$,  we have that
\begin{equation}\label{continuityxc}
 \partial_t \delta x_c=-\frac{1}{\rho}\nabla\cdot\left[x_c(1-x_c)\rho\delta
{\bf w}\right] -\delta {\bf v}\cdot\nabla x_c-\frac{\delta\Gamma_n}{\rho} \,.
\end{equation}

The  linearized Euler equations for both the comoving and
countermoving velocities are given by
\begin{eqnarray}\label{rocket1}
\partial_t\delta v_i+2\epsilon_{ijk}\Omega^j\delta v^k+\frac{1}{\rho}
\nabla_i\delta
p-\frac{\delta\rho}{\rho^2}\nabla_ip+\nabla_i\delta\Phi=(1-\bar\epsilon)\frac{
\bar\Gamma_n}{\rho}\delta w_i \,,
\\ \label{rocket2}
\partial_t(1-\bar{\epsilon})\delta w_i+\nabla_i(\delta\beta)+
2\bar{B'}\epsilon_{ijk}\Omega_j \delta
w^k-2\bar{B}\epsilon_{ijk}\hat{\Omega}^j\epsilon^{klm}\Omega_l \delta
w_m=(1-\bar\epsilon)\frac{\bar\Gamma_n}{\rho_c}\delta w_i \,,
\end{eqnarray}
where here we have defined 
$\bar{\epsilon}
=\epsilon_c+\epsilon_n=\epsilon_n(1+\rho_n/\rho_c)=\epsilon_n/x_c$, and where 
\begin{equation}
 \delta\beta=\delta{\tilde \mu}_c-\delta{\tilde \mu}_n
\end{equation}
and
\begin{equation}
\bar B=B/x_c \ , \qquad \bar B'=1-B'/x_c \,.
\end{equation}

The hydrodynamical equations can be studied employing a perturbative expansion
of the various hydrodynamical variables in $\Omega$, the star rotation
frequency. Actually, the expansion is in the parameter
$\Omega /\Omega_K$, where $\Omega_K$ is the Kepler frequency of the star.   
For superfluid systems, this expansion is particularly convenient, as  one can
show that the complicated system of equations for the comoving and countermoving
degrees of freedom decouple as these variables depend on different powers of
$\Omega$.

In the study of the evolution equations we shall restrict to the case where $\bar \Gamma_n=0$ and therefore 
the rocket terms in Eqs.~(\ref{rocket1}) and (\ref{rocket2}) will be neglected. The only contribution to dissipation 
will arise from the mass creation rate in
Eq.~(\ref{continuityxc}), and we shall evaluate the corresponding damping timescale  employing the
energy integral approximation, see {\it e.g.} \cite{Haskell:2009fz}. 

For our study we consider some simplifying,  admittedly unrealistic,
assumptions. We neglect the deformation of the star due to  rotation, which
affects the hydrodynamical variables at order $\Omega^2$. We use the
Cowling approximation, that is, we  neglect perturbations of the
gravitational potential associated with the oscillations of the star.
As a further simplification, we also consider a model where the mass density
of the star is uniform. As emphasized in the Introduction, our goal is to  study
the impact of the rocket effect in the evolution of the r-modes, and we
leave for future studies a more realistic model of the star.

Oscillations of a fluid element of a  stars can be described by the Lagrangian
displacement vector $\bf \xi$, which can be decomposed into a sum of toroidal
and spheroidal components. Since  neutron stars can be described
employing the two fluid model, one defines  comoving and countermoving
displacements, respectively ${\bf \xi}_+$
and ${\bf \xi}_-$, by means of the equations 
\be
\label{velocities}
\delta {\bf v} = \partial_t {\bf \xi}_+ \propto \Omega \,{\bf \xi}_+ \,,  \qquad
\delta {\bf w} = \partial_t {\bf \xi}_- \propto \Omega \,{\bf \xi}_- \,. 
\ee 
These two displacements describe the
center of mass oscillation and the out of phase oscillation of the two fluids,
respectively. We then expand these quantities in terms of
toroidal and spheroidal components
\bea
\xi_+ &=&
r \sum_{l,m}\left(0,\frac{K_{lm}}{\sin\theta}\partial_{\phi},-K_{lm}\partial_{
\theta}
\right)Y_{lm}+ r \sum_{l,m}\left(S_{lm},Z_{lm}\partial_{\theta},  \frac { Z_
{lm}}{\sin\theta}\partial_{\phi}\right)Y_{lm} \,, \\
\xi_- &=& r
\sum_{l,m}\left(0,\frac{k_{lm}}{\sin\theta}\partial_{\phi},-k_{lm}\partial_{
\theta}
\right)Y_{lm}+ r \sum_{l,m}\left(s_{lm},z_{lm}\partial_{\theta},  \frac {z_
{lm}}{\sin\theta}\partial_{\phi}\right)Y_{lm} \,,
\eea
where $Y_{lm}$ are the spherical harmonics. The
fluctuations of the pressure and of the chemical  potential difference can be
written  respectively as
\bea
\delta p &=& \rho g r \sum_{l,m} \zeta_{lm} Y_{lm}   \,, \\
\delta \beta &=& g r \sum_{l,m} \tau_{lm} Y_{lm} \,,
\eea
where  $g = \Omega_0^2 r$ (with  $\Omega_0^2 =
GM/R^3$) fixes the scale of pressure and chemical potential fluctuations.
Notice that with these definitions, $ \tau_{lm}$ and $ \zeta_{lm}$ are
dimensionless. 

Since in a superfluid star one has two different kind of displacements, in
principle one can have   two different kind of r-mode
oscillations, one associated with the comoving dof and one associated
with the countermoving dof.
Actually, the  hydrodynamical variables defined above 
obey a complicated set of coupled differential equations, see
\cite{Haskell:2009fz}, with couplings between  comoving and countermoving
displacements.
However,
as shown in \cite{Haskell:2009fz}, at the leading order in $\Omega$, one finds
that the equation for the comoving displacement decouples and one can determine
an
analytic expression for the  standard r-mode oscillation. Regarding the mode
associated with the  countermoving dof, it turns out to be a general 
inertial mode. That is, it is not a mode dominated by the toroidal components.
However, for incompressible stars with uniform density one has that this
inertial mode turns into an r-mode. We shall restrict to this case and analyze
this r-mode oscillation in Sec. \ref{sec:superfluid}. We explicitly
consider the effect of the mutual friction in the equations of motion, the
reason is that in this way we can analyze the regime where the mutual
friction coefficients, $B$ and $B^\prime$, are large. Therefore our results will
explicitly  depend on the values of these parameters. 

\subsection{Standard r-mode oscillations}\label{sec:standard}

For the standard r-mode oscillations  one assumes that the comoving toroidal
displacement,  $K_{lm}$, is of order
unity,  while the spheroidal
comoving displacements are  of order  $\Omega^2$. All the
countermoving
displacements turn out to be of order $\Omega^2$ as well.  Since the standard 
r-mode oscillation is dominated by $K_{lm}$, 
it is very similar to the r-mode oscillation  in  normal
fluids~\cite{Andersson:2000mf}, and can be easily determined  after
imposing proper boundary conditions~\cite{Haskell:2009fz}.
To first order in the rotation frequency of the star, one has that the typical
frequency of the oscillations (measured in the corotating frame) is 
\be\label{frequency-standard}
\omega_r = \frac{2 m \Omega}{l(l+1)} \,.
\ee
In our analysis we restrict to analyze the    case $l=m=2$, which corresponds to
the most unstable r-mode.

Regarding the pressure perturbations, they are of order $\Omega^2$, whereas
$\delta \beta \propto \Omega^4$ \cite{Lindblom:1999wi,Haskell:2009fz}. The order
in $\Omega$ of the toroidal oscillations and of the pressure and
chemical potential fluctuations are reported in the first line of Table
I.

\begin{table}[ht!]
\label{table:orders}
\centering 
\begin{tabular}{|c|c|c|c|c|c|} 
\hline  
type of r-mode & $K_{lm}$ & $k_{lm}$ & $\zeta_{lm}$ &
$\tau_{lm}$ & $S_{lm}, s_{lm}, Z_{lm}, z_{lm} $ \\ [1.ex]
\hline
standard r-mode& ${\cal O}(\Omega^0)$ & ${\cal O}(\Omega^2)$ & ${\cal
O}(\Omega^2)$ & ${\cal O}(\Omega^4$)& ${\cal O}(\Omega^2$)\\ 
superfluid r-mode & ${\cal O}(\Omega^2)$ & ${\cal O}(\Omega^0)$ &
${\cal O}(\Omega^4)$ & ${\cal O}(\Omega^2)$ & ${\cal O}(\Omega^2$) \\ [1.ex]  
\hline
\end{tabular}
\caption{Order in $\Omega$ of the comoving and countermoving displacements, 
of the pressure fluctuation and of the  chemical potential fluctuation for
the standard r-mode oscillation and for the superfluid r-mode oscillation.}  
\end{table}

For the purpose of estimating the damping time scales associated to both mutual
friction and the  rocket effect, carried out in the accompanying paper 
\cite{rocket2:2010}, we have to determine the solutions for the countermoving dof. 
The equations governing the evolution of the various dynamical variables are reported 
in the Appendix \ref{sec-evolution}.
Assuming constant mass density and  hydrostatic equilibrium  we
find that $\tau_{l+1}$ obeys the following differential equation
\be\label{differential-tau}
r^2 \tau_{l+1}^{\prime\prime} = (A_1+B_1-1) r \tau_{l+1}^\prime +(A_2 B_2 -A_1
B_1) \tau_{l+1} - A_2 B_4\, 
\frac{r^{l+1}}{R^2-r^2}\,,
\ee
where the prime indicates differentiation with respect to $r$ and the
coefficients
$A_i, B_i$ are  reported in Appendix \ref{sec-evolution}.
As shown there, the last term on the right hand side of this equation arises
because we have assumed that matter is  in hydrostatic 
equilibrium. The differential equation has  solution given by
\be
\tau_{l+1}(r) = f(r) + C_1 r^{n_1}+ C_2 r^{n_2} \,,
\ee
where $f(r)$ is the particular solution of the differential equation and 
where $C_1$ and $C_2$ are the coefficients of the homogeneous solution, to
be fixed by the boundary conditions. The exponents of the homogeneous solution
are given by
\be
n_{1,2} = \frac{A_1+B_1 \pm \sqrt{(A_1+B_1)^2+4(A_2 B_2 -A_1 B_1)}}{2} \,,
\ee
and it turns out that  $n_2$ is negative, 
meaning  that in order to avoid divergences at $r=0$, it must be  $C_2=0$. It is
interesting to note that for
vanishing mutual friction one has that $n_1=l-1$ and $n_2=-(l+4)$. 
As a second boundary condition we assume that the chemical
potential difference vanishes at the surface of the star, that is
$\delta\beta(R) = 0$. 

For completeness we report the equation obeyed by the radial component of  the
countermoving spheroidal displacement, which is given by 
\be
\xi_-^r = \frac{r^2 \tau_{l+1}^\prime}{A_2} - \frac{A_1 r \tau_{l+1}}{A_2}\,.
\ee

\subsection{Superfluid r-mode oscillations}\label{sec:superfluid}
Assuming that  $k_{lm}$ is of order
unity one finds that   the spheroidal
countermoving displacements are of order  $\Omega^2$. 
The driving force on the countermoving displacement is the chemical potential
difference which turns out to be of order $\Omega^2$.
The order of the toroidal  comoving displacement 
depends on the compressibility of the fluid. For a compressible fluid it is of
order $\Omega^0$, while for an incompressible fluid it is of order $\Omega^
2$. The reason can be traced back to the
fact that comoving oscillations are
driven by pressure oscillations and it turns out that $K_{lm} \propto \Omega^{-2}
\zeta$.  For compressible fluids the
pressure oscillations are proportional to chemical potential oscillations and
therefore   $\zeta \propto \Omega^2$ and thus $K_{lm}$ must be of order unity. 
Moreover, for this kind of mode, comoving spheroidal displacements turn out to
be of the same order in $\Omega$ of comoving toroidal displacements, meaning
that for a compressible fluid  this oscillation is a generic inertial
mode and not an r-mode. Since for a compressible fluid various components of the
displacements are of the same order in $\Omega$, one has to solve a system of
coupled differential equations. 

The situation is much easily tractable  for incompressible fluids. In this case
one can assume that spheroidal
oscillations are of order ${\cal O}(\Omega^2)$ and then 
toroidal oscillations turn out to be of the same order. The
order in $\Omega$ of the various displacements and of the pressure
and chemical potential fluctuations for incompressible matter  are reported in
Table I. We shall
restrict the analysis to the case of incompressible fluids, where the comoving
and countermoving dof decouple, with
the superfluid r-mode oscillation  dominated by the toroidal displacement
$k_{lm}$.  To first order in the rotation frequency of the star and to first
order in the entrainment parameter,  the typical frequency of the superfluid
r-mode oscillation (measured in the corotating frame) is 
\be\label{frequency-superfluid}
\omega_r = \frac{2 m \Omega}{l(l+1)}(1+\bar \epsilon) \,.
\ee
As for the standard r-mode, we restrict to analyze the    case $l=m=2$, which
corresponds to the most unstable r-mode. Moreover we consider only small values
of the entrainment. 

The analysis of the the superfluid r-mode oscillation is very similar to the
one we have performed for the standard r-mode oscillations, with the roles of
$K_{lm}$ and of the pressure oscillations  interchanged with $k_{lm}$  and  the
chemical
potential oscillations. We find that for superfluid r-modes, $k_{lm}$ obeys the
same equation that $K_{lm}$ obeys  for standard r-modes, and
the chemical potential fluctuation obeys the same equation 
that  pressure fluctuation obey for  standard r-modes. Regarding the
pressure oscillation, $\zeta_{lm}$, one has to solve an equation analogous to
Eq.~(\ref{differential-tau}), but without the last term on the right hand side,
because we are now considering an incompressible fluid.
We find that  
\be
\zeta_{l+1} = C_1  r^{s_1}+ C_2 r^{s_2} \,,
\ee
where $s_{1,2}$ depend on the parameters of the model. One of the
two coefficients is always negative, and therefore in order to avoid the
divergence at the origin, we have that 
\be
\zeta_{l+1} = C  r^{s} \,.
\ee
We fix  $C$ by demanding that the  comoving toroidal
displacement, $K_{lm}$, is properly normalized, as in Ref.~\cite{Lee:2002fp}.

\section{Conclusion}\label{sec-conclusion}
Superfluid neutron stars are characterized by various oscillation modes.  
Of particular interest are r-mode oscillations, because in the absence of efficient 
dissipative mechanisms they  lead to a rapid spin-down of the compact star. 
The reason is that r-mode oscillations couple to  gravitational-waves, and 
the  emission of gravitational waves (which spins down the star)  makes these 
oscillations larger. This unstable mechanism can however be damped by dissipative forces, 
which tend to reduce the amplitude of r-mode oscillations.  Indeed, if the characteristic  
timescale of the dissipative force is comparable with the timescale associated to the 
gravitational wave emission, the r-mode oscillation becomes stable, meaning that the 
compact star does not quickly spin-down by gravitational wave emission.

We have derived the perturbed hydrodynamical equations for two different r-mode 
oscillations in presence of the rocket effect, that is, in the presence of processes 
that change the number of protons, neutrons and electrons. The  two different r-mode 
oscillations considered are the ``standard r-mode oscillation", which is a  predominantly  toroidal comoving
displacement of the two superfluid and normal components, and  the  ``superfluid r-mode oscillation", which
is associated to toroidal countermoving displacements of the two fluids.  In realistic neutron
stars these two modes are coupled, however, in the limit of small rotation frequency and 
assuming that the star has a uniform mass density and is incompressible, they decouple. 
For both kind of oscillations we have determined the linearized Euler equations  and found 
that for both the standard r-mode oscillation and the superfluid r-mode oscillation the 
rocket effect leads to the appearance of additional dissipative terms in the perturbed 
hydrodynamical equations. These terms might give a relevant contribution to  the  energy 
dissipation of the oscillations, with a damping timescale comparable to the one associated 
to gravitational wave emission. The numerical evaluation of the corresponding damping 
timescale and the comparison with those deriving from other
dissipative mechanisms is performed in the accompanying paper \cite{rocket2:2010}.

\appendix

\section{Evolution equations}\label{sec-evolution}
\subsection{Standard r-mode}
We derive the evolution equations for the standard r-modes, 
assuming uniform mass density of the star.   For a
star with uniform mass density one can impose hydrostatic equilibrium obtaining 
\be\label{EoS}
P(r) = G \frac{2\pi}{3}  (R^2-r^2) \rho^2  \,,
\ee
where we have assumed that the pressure vanishes at the surface of the star.
Here $R= 10$ km is the radius of the star and we shall consider a mass density
$\rho = 2.5 \rho_0$, where $\rho_0$ is the saturation density of nuclear
matter. 
With these values we obtain that the mass of the star is $M \simeq 1.47
M_\odot$, where $M_\odot$ is the mass of the sun.
In this case we have that the pressure and the various components of the
countermoving mode obey the following set of equations~\cite{Haskell:2009fz}  
\bea\label{klm}
k_{lm} &=& a\, s_{l+1} + b\, z_{l+1}  \\ 
\label{tau}
\tau_{l+1} &=& c\, k_{lm} + d\, z_{l+1} +e\, s_{l+1} \\ 
\label{tau-diff}
r \frac{d \tau_{l+1} }{d r} &=& -2 \tau_{l+1} - f\, k_{lm} - g\,
z_{l+1} + h\, s_{l+1} \\
\label{s-diff}
r \frac{d s_{l+1} }{d r} &=& - 3 s_{l+1} -
\frac{V}{\Gamma}\frac{1}{1-x_p} \zeta_{l+1} + p\,
z_{l+1} \,,
\eea
where the various coefficients have been derived in Ref.~\cite{Haskell:2009fz}
and for  $l=m$ are given by 
\begin{center}
\[\begin{array}{rlrl}
a &= \frac{ \bar B - i  \bar B^{\prime}}{(1-\bar \epsilon) -  \bar
B^{\prime} - i {\bar B}} \frac{1}{\sqrt{2 l +3}}\,&
\qquad b &= (l+2) a\, \\
c &=  l \omega^2  ( \bar B -i \bar B^\prime ) \frac{1}{\sqrt{2 l +3}}\,&
\qquad d &= \frac{\omega^2}{l+2} \left((l+2) (1-\bar \epsilon) - l   \bar B^{\prime} -
 i   \bar B\frac{2 + 11 l + 8 l + 2 l^2}{5 + 2 l}\right)\, \\
e &= - \frac{\omega^2 }{l+2} \left(l \bar B^\prime - i
B \frac{2-l}{5+2 l}\right)\,&
\qquad f &= -(l+1) c \,\\
g &= -p\, e \,&
\qquad h &=  \omega^2 \left((1-\bar \epsilon) -2  i \bar B
\frac{(l+1)^2}{5+2 l}\right) \,\\
p &= (l+1)(l+2) \,
\end{array}\]
\end{center}
where  $\omega=\sigma/\Omega_0$, with
$\Omega_0 = \sqrt{GM/R^3}$.  In Eq.~(\ref{s-diff}) we have that 
\be\label{definitions}
V= \frac{g r \rho}{P} \qquad {\rm and} \qquad \Gamma = \frac{d \log P}{d \log
\rho} \,,
\ee
which depend on the equation of state. 
Since we assume hydrostatic equilibrium we have from Eq.~(\ref{EoS}) that 
\be
V = \frac{2 r^2}{R^2-r^2}\qquad {\rm and} \qquad \Gamma=2 \,.
\ee 
Notice that we shall assume that both the background and the perturbations will obey the same equation of state, 
therefore the coefficient $ \Gamma$ determined above, characterizes both the background and the perturbation.

Finally, according with Ref.~\cite{Haskell:2009fz}, we have that the pressure
fluctuations are given by 
\be
\zeta_{l+1} = 2\frac{\omega \tilde \omega }{\sqrt{2l + 3}} \frac{l}{l+1}
\frac{r^{l-1}}{R^{l-1}}  \,,
\ee
where $\tilde \omega = \Omega/\Omega_0$ . Upon substituting the expressions above in the equations~
(\ref{klm}), (\ref{tau}), (\ref{tau-diff}) and (\ref{s-diff}) and expressing
$k_{lm}$ and $z_{l+1}$ in terms
of
$s_{l+1}$ and $\tau_{l+1}$ we have two coupled  differential equations for
$s_{l+1}$ and $\tau_{l+1}$. These equations  can be written as a second order
differential equation 
\be
r^2 \tau_{l+1}^{\prime\prime} = (A_1+B_1-1) r \tau_{l+1}^\prime +(A_2 B_2 -A_1
B_1) \tau_{l+1} - A_2 B_4 
\frac{r^{l+1}}{R^2-r^2}\,,
\ee
where
\begin{center}
\[\begin{array}{rlrl}
A_1 &=\, -2 - \frac{fb + g}{cb + d} \,,&\qquad 
A_2 &=\, - f \frac{ad -be}{cb + d} + g \frac{ac +e}{cb + d} + h\,,  \\
B_1 &=\, - 3 - p \frac{ac +e}{cb + d}\,, &\qquad 
B_2 &=\, \frac{p}{cb + d}\,, \\
B_4 &=\,  \frac{z}{1-x_p} \,.
\end{array}\]
\end{center}

The analysis of the superfluid r-modes is analogous to the one we have done for
the standard r-modes. However, in order to have an r-mode oscillation and not a
generic inertial mode, one has to assume that the fluid is incompressible
\cite{Haskell:2009fz}.   In this case $\Gamma \to \infty$ and the differential
equations one has to solve are simpler.

\subsection{Superfluid r-mode}

In the case of the superfluid r-mode we assume that the fluid is incompressible, 
that means $\Gamma\rightarrow\infty$. Therefore, in this case we are left with 
the following equations for the countermoving degree of freedom:
\bea
r\frac{d\tau_{l+1}}{dr} &=& (l-1)\tau_{l+1}\\
k_{lm} &=& -\frac{l+1}{2i\omega\tilde\omega Q_{l+1}(l\bar B'+im\bar B)}\tau_{l+1}
\eea
while for the comoving degree of freedom
\bea
r\frac{dS_{l+1}}{dr} &=& -3S_{l+1} + aZ_{l+1}\\
r\frac{d\zeta_{l+1}}{dr} &=& -2\zeta_{l+1} + bS_{l+1} + cZ_{l+1} + dK_{lm}\\
\zeta_{l+1} &=& eK_{lm} + fZ_{l+1} + gS_{l+1}\\
K_{lm} &=& hS_{l+1} + jZ_{l+1}
\eea
where we derived similar coefficients as in Ref.~\cite{Haskell:2009fz} and, for $l=m$, are given by
\begin{center}
\[\begin{array}{rlrl}
a &= (l+1)(l+2)&\qquad
b &= \omega^2 \\
c &= -2l\omega\tilde\omega&\qquad 
d &= -2i\omega\tilde\omega lQ_{l+1} \\
e &= \frac{d}{l+1}&\qquad 
f &= \omega\left[\omega-\frac{2\tilde\omega}{(l+1)(l+2)}\right] \\
g &= -\frac{2}{l+1} &\qquad 
h &= -\frac{i\omega_0}{\omega-\omega_0}Q_{l+1}\\
j &= (l+2)h 
\end{array}\]
\end{center}

Now, by expressing $K_{lm}$ and $Z_{l+1}$ in terms of $S_{l+1}$ and $\zeta_{l+1}$ 
we are left with two differential equations for the latter variables, that can also 
be rearranged as a second order differential equation for $S_{l+1}$
\be
r^2 S_{l+1}^{\prime\prime} = (A_1+B_1-1) r S_{l+1}^\prime +(A_2 B_2 -A_1
B_1) S_{l+1},
\ee
where
\begin{center}
\[\begin{array}{rlrl}
A_1 &=\, -3 - a\frac{eh + g}{ej + f} \,,&\qquad 
A_2 &=\, - c \frac{eh +g}{ej + f} -d \frac{gh-fh}{ej + f} + b\,,  \\
B_1 &=\, - 2 + \frac{c}{ej + f}+ \frac{dj}{ej + f}\,, &\qquad
B_2 &=\, \frac{a}{ej + f}\,.
\end{array}\]
\end{center}

\begin{acknowledgments}
This work has been supported in part by  the INFN-MICINN grant 
with reference number FPA2008-03918-E. The work of CM has been supported  by the
Spanish  grant FPA2007-60275 and FPA2010-16963. The work of MM has been
supported by the Centro Nacional de F\'isica de Part\'iculas, Astropart\`iculas
y Nuclear (CPAN) and by the Ministerio de Educaci\'on y Ciencia (MEC) under
grant  FPA2007-66665 and 2009SGR502.   
\end{acknowledgments}

\end{document}